# Empowering the Teaching and Learning of Geometry in Basic Education by Combining Extended Reality and Machine Learning

Carlos R. Cunha[1][0000-0003-3085-1562], André Moreira[2][0000-0002-6253-6615], Sílvia Coelho[3][0009-0006-4333-1151], Vítor Mendonça[2][0000-0001-7020-8235] and João Pedro Gomes[2][0000-0001-9308-0027]

[1] Research Centre in Digitalization and Intelligent Robotics (CeDRI), Instituto Politécnico de Bragança, Campus de Santa Apolónia, 5300-253, Bragança, Portugal
`crc@ipb.pt`

[2] Instituto Politécnico de Bragança, Campus de Santa Apolónia, 5300-253, Bragança, Portugal
`andre.moreira@ipb.pt, mendonca@ipb.pt, jpgomes@ipb.pt`

[3] Escola Básica 2,3 Dr. Francisco Sanches, Rua do Taxa, 4710-448, Braga, Portugal
`silvia.rc.coelho@sapo.pt`

**Abstract.** Technology has helped to innovate in the teaching-learning process. Today's students are more demanding actors in the mechanisms available to them to learn, experiment and develop critical thinking. The area of mathematics has successively demonstrated the existence of learning difficulties, whether due to students' lack of motivation, low abstraction capacity and lack of new tools for teachers to innovate in the classroom and outside it. If it is true that digitalization has entered schools, it often follows a process of digital replication of approaches and materials that were previously only available on physical media. This work focuses on the use of Extended Realities for teaching mathematics, and very particularly in teaching geometry, where a conceptual model is proposed that combines the use of Extended Reality and Machine Learning. The proposed model was subject to prototyping, which is presented as a form of laboratory validation of the proposed model and presents itself as a contribution to innovate in the way the geometry teaching-learning process is developed and through the ability to obtain useful insights for teachers and students throughout the process.

## 1  Introduction

Ensuring access to technology is a key step for schools to convert to digital. However, for this conversion process to be successful, it is vital that the focus is not on technology but on understanding how technology can enable teaching and learning in an effective and inclusive way [1], [2].

Nowadays, educational projects include the physical manual as well as digital materials, such as the manual in digital format, question banks, worksheets and tests.

It would be expected that the teaching and learning of Mathematics and in particular Geometry would be greatly strengthened and, consequently, school results would improve. However, many times this barely happens, because often, the so-called technological support materials used are just a digitization of what exists in paper format – solving exercises and problems by only "writing on the computer", using the same paradigms, stops being interesting after a while.

Also, a strong obstacle shared by many mathematics teachers is the lack of time to work/deepen each topic to be taught and, therefore, also Geometry. They are faced with a lack of interest on the part of students in showing attention and concentration in classes, in being focused on the subjects covered by the teacher, on the tasks to be carried out and, with a lack of continuity of work at home. According to [3], the lack of adequate educational methods is becoming increasingly evident, leading to a decrease in students' interest in learning and critical thinking.

Without resolving these basic problems, there is no prospect of improvements in terms of knowledge acquisition and application. For instance, many students do not read the statements properly and give up thinking simply because a geometric figure is in a position that the student considers strange, such as the position of a solid. Very often they are unable to apply the Pythagorean theorem in space, as they reveal difficulties at the level of abstraction. A study on the difficulties of applying the Pythagorean theorem can be found in [4]

On the one hand, there has to be a tripartite commitment between School, Parents and Students to boost the teaching/learning process. If one fails, the entire teaching/learning process is compromised. According to [5], ensuring communication, and supporting tools, among the different actors present in the school ecosystem is an essential issue, and the use of appropriate technology can bring the different actors closer together.

Students like to understand the application of Mathematics in everyday life and in the world of work. They need to carry out project work, to work on something that excites them – using the right tools can give a strong boost to this entire process.

Technology plays a key role in shaping the future of education [6]. New technologies have undergone significant advances, making it possible today to integrate them into the teaching/learning process in a more valid way for educational practices in teaching. There is a wide variety of innovative equipment, programs and digital games that offer special meaning in the construction of knowledge and are appealing. Well chosen, used appropriately and sparingly, planned according to the characteristics/profile of the targeted students and taking the time these students need to be agents who build their knowledge through logical/deductive reasoning, they are certainly a strong plus – motivating the student to learn, without compromising the basic principles of each subject.

Dynamic Geometry environments can promote the teaching and learning of Geometry. It can help overcome difficulties or even avoid some of them. In fact, according to [7], the use of Dynamic Geometry Systems (DGS) has attracted special attention within mathematics education. This way of approaching the study of Geometry allows students to carry out tasks of a more exploratory and investigative nature, creating an environment where those involved interact, discuss ideas, formulate conjectures, solve problems and, consequently, obtain generalizations with relative ease, being able to justify the obtained results. When they become protagonists of their own learning, students develop characteristics that contribute to aspects of their social, personal and professional lives.

One of the advantages of using new technologies in teaching Geometry is having dynamic access to complex figures that would otherwise be difficult to view from another perspective – knowledge becomes quick, easy, interactive and accompanied by logical reasoning. In a study carried out by [8], the results indicated that the implementation of new technologies such as Virtual Reality (VR) and Augmented Reality (AR) improve interactivity and students' interest in teaching/learning mathematics, contributing to learning and more efficient understanding of various mathematical concepts, such as the study of solids, when compared to the use of traditional teaching methods.

Empowering teachers and students, through the use of appropriate technologies, is vital for a change in the teaching/learning paradigm. Since mathematics is often perceived as something abstract and difficult to visualize its laws and dynamics, as well as its applicability, it is important to provide mechanisms for rendering these same laws and applications.

However, it is equally important that when using technology, we are able to understand the way in which students conduct their learning process, being able to personalize their experience and drawing useful insights for the teacher to respond to the needs of each student. In this way, teaching actors will be able to maintain high levels of motivation, generate greater capacity for autonomous students-learning and student-exploration, in an immersive environment that enable critical thinking.

According to [9], immersive media presents benefits in terms of increasing motivation and expanding traditional teaching practices, involving students in different ways.

This article proposes a model based on the use of Extended Reality (XR) – to support the immersiveness and objectification of the not always objective dynamics of mathematics. And, Machine Learning (ML) – to personalize the student XR experience and generate knowledge that will help the teacher to better understand and frame the students' difficulties. The potential of ML is studied by [10], where the role that this technology can have in the ability to predict student performance is highlighted.

The proposed approach – a conceptual model that combines XR and ML – is a contribution to supporting an immersive vision based on computational intelligence, of the teaching/learning process. It is also presented a prototyping of the model presented for the teaching of mathematics and in particular in the teaching of geometry taught in the context of basic education.

## 2 Literature Background on The Use of Extended Reality and Machine Learning on Education

The use of technology to innovate in the teaching-learning process is nothing new. In fact, according to [11], integrating information technologies as a tool capable of assisting in education is mandatory, especially in an era in which there is a notable transition in the teaching-learning process.

In a study carried out by [12], which analyzes the role of reality-enhancing technologies for the teaching and learning of mathematics, some conclusions point to the need to encourage the use of these technologies by both students and teachers and, that teachers need to know more than just about the tools used, but also how they can teach through it. Also, in a study carried out by [13], which analyzes, using a systematic review, the different approaches to learning with Immersive Virtual Reality (IVR), educational researchers are urged to consider existing approaches and rethink the process used to design IVR-based learning tasks to achieve their pedagogical goals. However, according to [14], IVR has the potential to transform traditional classrooms into immersive virtual reality scenarios that are effective in the learning process and for research purposes.

With regard to geometry teaching, the use of AR is described by several authors. According to [15], over the past decade, the use of AR has proliferated in the education sector. However, the number of articles that systematically reviewed the research trends in the implementation of AR for learning mathematics was reduced.

In [16], a structure of a learning environment system based on AR, supported by mobile devices, is presented. This system allows students to get assistance as well as tips to solve problems in the context of geometry. In [17], is proposed a design, implementation and evaluation of an AR-based geometry learning application where interactions are based on hand gestures. However, the model does not include the use of ML to understand student difficulties and/or personalize teaching-learning experiences.

Also, in [18], AR was used to develop material as a support mechanism for teaching and learning mathematics and examining its effectiveness. However, despite the improved students' learning outcome, it is stated that the development of AR material presents some difficulties, being necessary to solve technical problems, proceed with the improvement of certain features, and be able to provide more clear instructions to users. In this work, the use of AR and ML was also not combined.

In [19], is presented a work where AR was used to assist learning achievement, generate motivation and creativity for children of low-grade in primary school. The results of this investigation showed a positive impact on the degree of student satisfaction in their teaching-learning process, an increase in students' motivation to learn less popular subjects (e.g., such as geometry), and an improvement in creative thinking.

In a study carried out by [20], which analyzed the potential of AR for Teaching Mathematics, more specifically for teaching Vector Geometry, the use of AR in mathematics classes was considered beneficial and fun by the teacher and students. The use of AR has been shown to be an aid in spatial imagination, something that can be difficult to accomplish when using only traditional 2D material.

Although there are several studies and applications of XR, namely AR and VR in the field of teaching, the approaches used do not combine the potential of XR with ML, reducing the ability to generate intelligence each time a student uses an XR application in their learning process. In this way, is not taken advantage of the full potential return that would come from knowing the study patterns, failures and approaches that students have when solving or experimenting educational materials, built with XR. Combining

ML with XR will allow students to better understand their failures and, for teachers, it will provide a personalized teaching management tool, with a positive impact on the readjustment of methodologies and pedagogy used in the classroom.

ML can help understand the student and support the provision of adapted and personalized content, materializing an efficient recommendation system in the learning environment, considering student behaviors and preferences when recommending various learning materials - a system that adapts to the needs and student learning skills [21].

## 3    Proposed Conceptual Model

This section presents a conceptual model (see Fig. 1) based on the use of XR and ML and intended to be a contribution to innovating in the teaching-learning process. The model is applicable in any area of teaching; however, it is the target of prototyping for the teaching of mathematics and more specifically for the teaching of geometry in basic education.

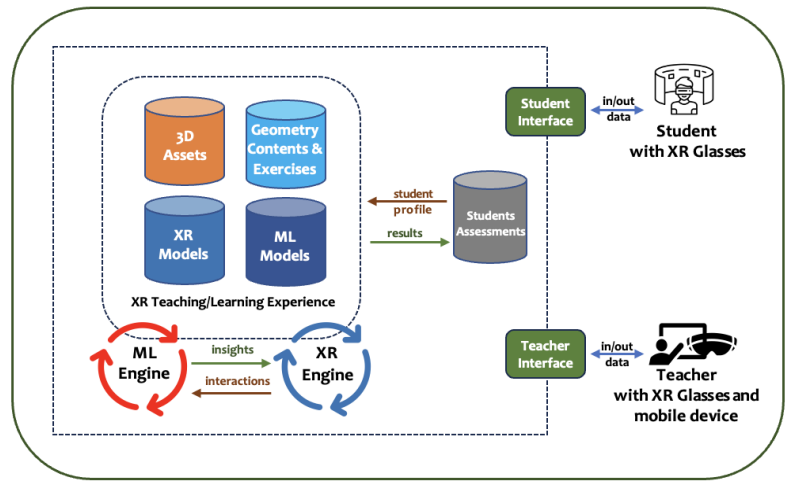

**Fig. 1.** Proposed Conceptual Model.

The proposed model has several components and actors that are presented below and where their main objectives and form of action are characterized:

**3D Assets Database:** Represents the repository of 3D assets that are used in XR applications.

**Geometry Contents & Exercises:** It represents the content associated with the subject that will be taught and/or the object of study to be developed inside or outside the classroom. This content will be combined with 3D assets to support XR applications.

**XR Models:** It represents the set of models/scenarios that will be personalized depending on the teacher's management in the context of teaching-learning in the classroom, homework or free study activities carried out by students.

**ML Models:** It represents a database that has been trained to recognize certain types of patterns obtained through data sets and algorithms that are used to weight and learn from that data (which are obtained from the data trail that is left by students when using XR applications)

**XR Engine**: Represents the module that extracts the actions carried out by student and learner, feeding the ML engine. It is capable of generating personalized XR scenarios based on input from the ML engine.

**ML Engine:** Represents the intelligence engine of the proposed model. It is capable of, analyzing the insights and actions carried out by students when carrying out activities, interacting with the XR Model generating new scenarios of personalized activities depending on the difficulties of each student, as well as, in real time in the generation of each new step of the activity (e.g., next question).

**Students Assessments:** represents the repository assessments as well as the results of these assessments by student. This is useful for teachers and also to understand students' difficulties.

**Student Interface:** Represents the students-view of the system. Students will use XR Glasses (e.g., HoloLens® 2), and all the f functionalities thar students can use (e.g., collaborative working in classroom, homework, free activities at home).

**Teacher Interface:** Represents the teacher-view of the system. Teachers will use XR Glasses (e.g., HoloLens® 2), and all the f functionalities thar students can use.

In order to exemplify some functionalities to be implemented, the following use case diagram (see Fig. 2) illustrates the activities involved in classroom learning.

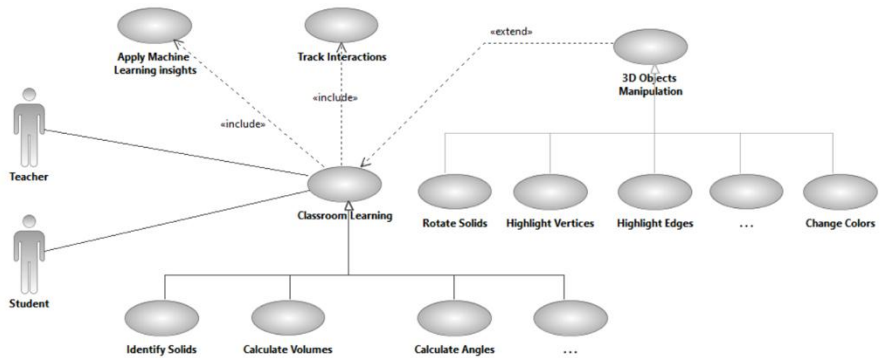

**Fig. 2.** Use Case – Classroom Learning.

Thus, as illustrated, it can be seen that the system constantly tracks the interactions that will serve as the basis for machine learning insights. In classroom learning, both the student and the teacher will be able to interact, allowing different types of learning (identification of solids, calculating areas and volumes, among others) and being able to manipulate objects in different ways (rotate solids, highlight vertices and edges, change colors, among others).

On the other hand, the system should allow the teacher to monitor the progress of students' learning, as represented in the following use case diagram (see Fig. 3).

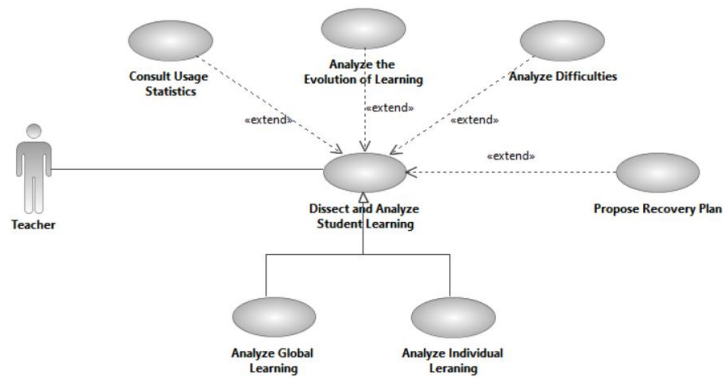

**Fig. 3.** Use Case – Analyze the evolution of learning.

Monitoring the evolution of learning, using statistical analysis and inference and understanding of the difficulties experienced by students, will allow the teacher to propose new challenges and recovery plans.

The conceptual model presented is a contribution to support innovation in the way we idealize and operationalize new teaching-learning paradigms, more capable of meeting students' motivations and contributing to their critical development. Likewise, it aims to contribute to the creation of collaborative environments in the classroom and help break barriers in the difficulty of abstraction associated with mathematics and especially in the teaching of Geometry, where the 2D view of materials on paper causes difficulties. Finally, it aims to help generate motivation in students to develop independent study work while providing valuable insights for the teacher.

## 4    Prototype Developed

This section presents the developed prototype based on the conceptual model presented in the previous section being referred the technologies used, the implemented features and planned ones as well. This is an ongoing work that aim the validation of the conceptual model proposed, and the development of a solution to be tested in real school classes context.

### 4.1 Technologies Used

When developing the prototype was considered the available equipment for Mixed Reality (MR) being this the HoloLens® 2, an Optical Head-Mounted Device (HMD) or Optical See-Through (OST) HMD [22], manufactured by Microsoft®, this device was released in 2019 and it works as a standalone HMD without the need of controllers or connection to external components, being this possible by featuring technologies like Hand Tracking, Eye Tracking and Spatial Mapping.

**Unity.** Also known as Unity3D or Unity Engine, is a Game Engine used primarily for game development, being highly recognized among the XR Developer duo to his low learning curve and ease-of-use when compared with others game engines like Unreal or CryEngine, the big community and support when developing XR applications as well for being a primary suggestion by most of the pioneers of this technology like Meta®, Microsoft® and VIVE®.

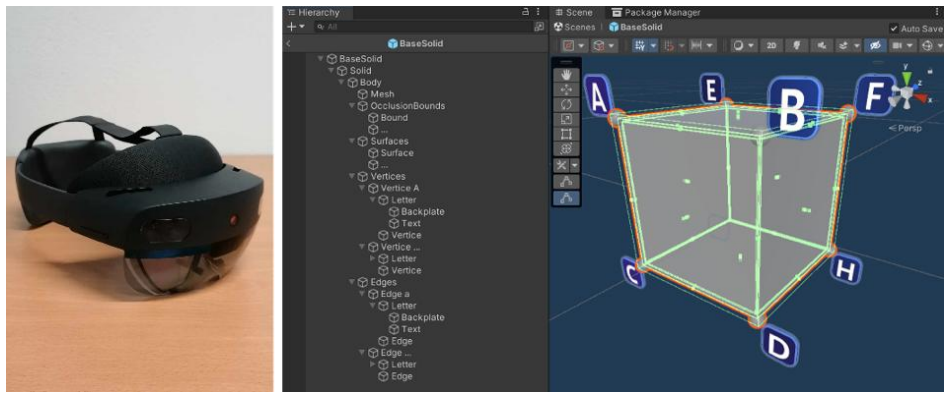

**Fig. 4.** HoloLens® 2 and Prototype 3D Primitive GameObject Architecture with Example in Unity.

**MRTK3.** The Mixed Reality Toolkit is a framework created by Microsoft® to provide the necessary tools to developers when creating MR applications to HoloLens® 2 in Unity, the current version is the MRTK3 and while available as public preview since June 2022, it was declared as generally available in September 2023. While focused on the HoloLens® 2, applications developed with MRTK2 and principally MRTK3 can also be built to be used in other devices that support MR like Magic Leap 2, Meta® Quest 2 and the most recent Meta® Quest 3.

### 4.2 Features Implemented

In this section will be described the implementation process of the three main concepts presents in the developed prototype.

**General Solid Architecture.** Geometry is the branch of mathematics concerned with the study of solids, lines, faces and vertices, as such these elements are the very base necessity to be replicated in MR being also necessary the possibility of treating each component as individual element. As such the solid, being a cube or a triangular prism for example, is mounted by defining each component individually (see Fig. 4), the vertices are defined by using 3D Vectors (Vector3) while the edges make use of a feature of Unity called Line Renderer being used the vertices to define the start and end of the line. However, for the definition of faces, as well to have the complete solid as individual component for some use cases and for more complex's solids, was necessary the use of primitives that Unity does not provides, being acquired the asset Deluxe Primitives by Reactorcore Games and Ultimate Procedural Primitives by KANIYONIKA from the Unity Asset Store.

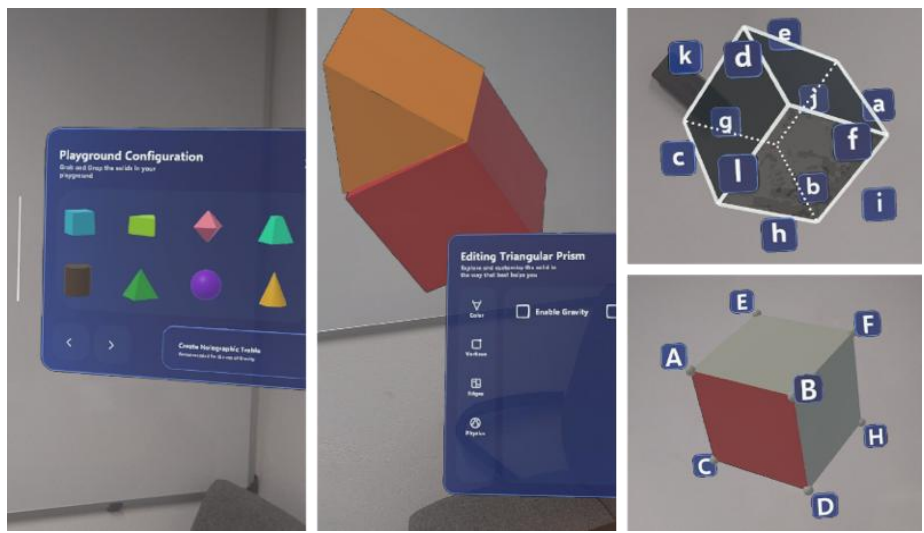

**Fig. 5.** 3D Primitives Creation screen and Solid Personalization Use-Cases

**Solid Personalization.** Consists in empowering the user, being a teacher or student, by enabling him to modify the solid in a way to better highlight a concept (see Fig. 5).

*Change Color.* Enables the user to highlight specific components of the solid or even differentiate them being a basic concept present both in school manuals or when used physical solid representations, but without limitations or constraints to presets.

*Edit Vertices.* Allows the user to modify the representation of the vertices, while they are defined through simple 3D points, the user can choose to display visual markers (spheres), or letters as well display only specific ones.

*Edit Edges.* Provides a group of possibilities such as display only the edges while hiding the solid faces as well display letters. The user can also choose to display only the outline of the solid or enable invisible edges.

*Physics.* Allows changing solid characteristics such as how it interacts with gravity, and mass, as well with other solids, which can be used for explaining geometric concepts such as cube number.

**Playground Configuration.** Refers to the basic ability of the user to create more solids at his own choice as to manipulate them both in terms of position, scale, and rotation in the digital world, as well to create composite solids. Depending on how the solid physics is configured those same solids can collide between them or be inserted inside each other to assist for example at calculating volumes.

### 4.3 Ongoing and Future Development

Aside from further improvements, based on the necessities noted during the initial tests, like the possibility of personalizing the surfaces to for example only being visible a set of them, the possibility of creating 2D primitives such as triangles instead of only 3D or clipping those primitives.

Some more advanced features are the possibility of streaming the user view to a desktop screen using Mixed Reality Capture (MRC) and WebRTC, enabling scenarios where the teacher may present concepts during classes while projecting the 3D solids for the class. Another feature is the development of shared experiences in a way that two or more users of the MR HMD may interact with the same 3D world similarly to Multiplayer Games by using technologies like Photon Unity Networking 2 (PUN 2).

Finally, the ML engine is being implemented and will be improved after experimentation in a real scenario (i.e., some school classes) and, defining the objectives that teachers consider most relevant to help personalize the proposed exercises and that will help to understand the reasons that tend to characterize the difficulties in understanding and learning, of each student.

## 5 Conclusion

Technology has supported and encouraged the introduction of innovation in education, promoting new methodologies inside and outside the classroom, as well as innovating the concepts of school manuals and study support manuals.

Among the technologies that are most promising, XR can enhance essential factors in the teaching-learning process, such as increased levels of motivation, increased capacity for abstraction and development of critical thinking.

Likewise, the use of XR to create cooperative teaching-learning environments allows the construction of a new paradigm of interaction between teacher and students, especially in the classroom.

The combination of XR with ML raises the potential of XR, allowing the obtaining of insights that can be decisive in understanding, on the one hand, students' difficulties in a personalized way and, on the other, the definition of more appropriate strategies inside and outside the classroom.

This paper proposes a conceptual model based on the use of XR and ML, as a contribution to the definition of new teaching-learning strategies. In order to validate the model, a prototype was developed using MR for teaching Geometry. The prototype is still under development for full validation of the model. In particular, its ML component, which intends to be implemented after experimentation in the classroom and the subsequent validation of its MR approach and feedback from teachers and students.

## Acknowledgements

Blind funding acknowledgements for peer review purposes.